# The Olbers Conjecture, Revisited*

**Revised and corrected for a poster presentation at Stanford Linear Accelerator Center Summer Session, SSI-31, August 2003**

By John Michael Williams

`jwill@AstraGate.net`

2003-07-05

* Available at `http://arXiv.org/pdf/physics/0104067`

## Abstract

The Olbers conjecture, that under reasonable assumptions, light from the stars should sum at the Earth to make the sky bright at night, has been a subject of study since the early 19th century. It has been incorporated into some of modern cosmology. To complement Olbers's conjecture, we suggest a new calculation modelled as a projecture, in the form of an imaginary star probe. We find that there are not enough of stars. We also confirm Olbers's reasoning analytically.

Note: The SSI-31 slides are numbered and presented at the end of the text; the slides include some results omitted from the text.

## Introduction

Heinrich Wilhelm Olbers (1758 - 1840), in 1823 presented an argument, many times repeated in many ways, by which it would seem that the night sky should be bright, not dark. An earlier conjecture on the same theme is attributed to Kepler and later de Chesaux, as pointed out in an historical review by [Newton, 2001].

All forms of Olbers's argument go like this: We assume the universe at least out to a very great distance to be populated with some average density of stars. Drawing a set of equally-spaced concentric spherical shells around the Earth, each $n$-th shell will contain a number of stars proportional on the average to its volume. The volume of each shell will increase as the square of its average radius; the angle on Earth subtended by a star in that shell will decrease as the square of that shell's average radius. The two squares cancel, so, for all shells beyond some reasonably great distance from Earth, the amount of star light contributed to Earth by each shell roughly will be equal to that of any other shell. But, the number of shells must be very great if not infinite; therefore, an observer on Earth should see the sum of all shells, a solid wall of stars, and so the sky should be very bright at night.

On thinking this through, one realizes that whereas a star can block a line of sight to empty, starless patches of sky, the reverse does not hold: Empty sky can not block a star. Therefore, the asymmetry seems to favor light over dark, strengthening the conjecture.

Notice that this result is not logically self-contradicting; so, strictly speaking, it is not a paradox. It is, rather, an apparently valid physical assertion which obviously can not be true. A paradox would be self contained and have no physical reference to fact: For example, the liar's paradox: "I never tell the truth". There is no meaningful physics involved in a paradox.

Because Olbers's logic is self-consistent, the present author will refer to it as the Olbers *Conjecture* as to the darkness of the sky, rather than as the "Olbers Paradox".

Edgar Allen Poe was familiar with the Conjecture and is said to have speculated that the newly measured finite speed of light was responsible:  Light from the farthest shells had not yet had time to reach Earth.   The general explanation accepted by astrophysicists is that the age of the universe, its finite size, and the density of the stars conspire to render the Olbers Conjecture ineffective.

One relativity web site [Chase, 1993] lists these explanations (reworded):

- There's too much cosmic dust
- The number of stars is too small
- The stars are not uniform enough; they block one another
- The expanding universe red-shifts distant stars into obscurity
- The age of the universe has not allowed time for all the light to reach us (Poe's argument).

A posting [Weiner, 1998] refuting an analysis by Marilyn Vos Savant uses a galaxy atlas to compute that the number of stars actually is too small.

Leaning on the age of the universe are more general computations by [Imamura, 2000], and some others by [Arpino & Scardigli, 2002].

However, with enough stars one can illuminate anything, so the age or the size arguments, while somewhat interdependent on each other and on the star density, do not necessarily imply each other.

Olbers's Conjecture seems to hang around forever in the sky.   Almost everyone introduced to it seems to ponder it a while, see that it seems reasonable, fail to find an error, and then do nothing much about it.

I would like first to construct a new but similar conjecture which should shed some new light on Olbers's and then provide yet another simple analytic confirmation of Olbers's own argument.

## A Telescopic Projecture

The upper half of Figure 1 shows an observation station on Earth equipped with a device which telescopes into space.  The device consists of a circular cylinder one parsec in diameter which may be extended into space indefinitely, pushing out the end cap on the right at close to the speed of light.   The distant end-cap surface of the cylinder is opaque, and the side walls exclude light except that originating within the cylinder. Stars may drift in and out of the cylinder; those within have all their emission in the direction of Earth captured and conveyed to Earth in either of two ways: (*a*) parallel modes as though by a huge bundle of optical fibers; or, (*b*) simple direct, square-law radiation (the operator flips a switch to choose), with walls of the cylinder perfectly absorbing.  An observer either may measure the total light gathered by the device at point *A*, or may look directly into the cylinder at point *B*.

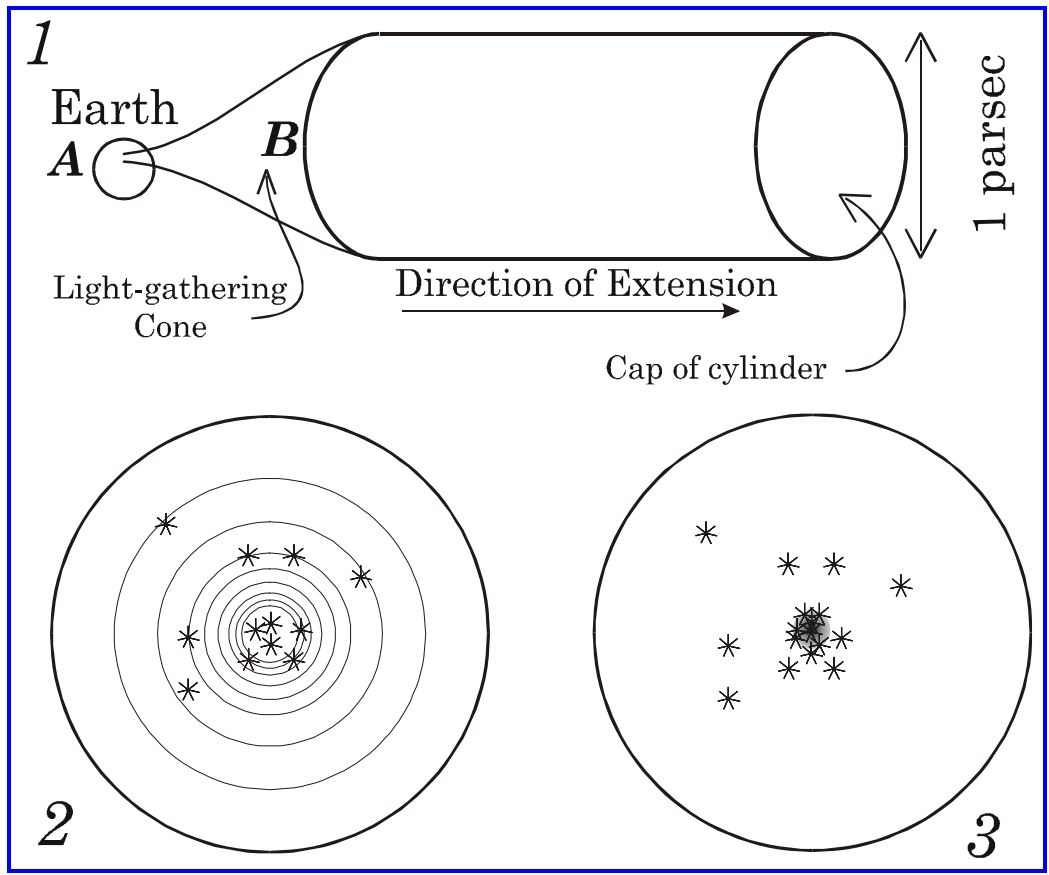

**Figure 1. (1) Telescopic mechanism to gather starlight. (2) View from *B* after operation for about 2000 years. The concentric field markers each represent about 100 parsecs. (3) view from *B* after an indefinite operating period (field markers omitted). Views 2 and 3 are not to scale and are magnified by some factor around 100x.**

My projecture is that if I wait long enough, I can use the fiber-optic mode to gather energy equivalent to the emission from the surface of a star 1 parsec in diameter. So, assuming the stars all are about uniformly distributed in space, say, an average of 1 per 10 cubic parsecs (an uncorrected, greatly overestimated density), and that each averages $10^6$ km in diameter (about $8 \cdot 10^{11}$ km$^2$ projected area), I wait. My telescopic device extends itself by about 3 parsecs per decade, or 0.24 cubic parsecs per year, and thus it encloses an average of about 2.4 new stars per century. So, I see about 1.2 new stars per century.

After about 2000 years, the cylinder is almost 2000 light-years long, and light from maybe 1000 years past is available. So, as shown in Figure 1, part 2, by direct viewing, I see perhaps a dozen stars. Notice that postulating an extensible telescopic device versus assuming immediate availability of starlight only makes a difference of a factor of two. Even after several thousand years, all I see by direct view is as shown in Figure 1, part 3. The reason is that the field angle subtended by the cap has become vanishingly small, and all the numerous, distant stars are bunched up almost to a point.   Also, because no star comes even close to filling a parsec, the average light collected at *A* in Figure 1 will be extremely weak.  Nevertheless, each additional period of time of operation adds on average the same new number of stars.

Switching to the fiber-optic view, though, how long will I have to wait before I have achieved my goal?

A parsec is about 3.26 light-years or $3.26 \cdot (3 \cdot 10^8) \cdot (24 \cdot 60 \cdot 60 \cdot 365)$ meters, or about $3 \cdot 10^{13}$ km.  So, the telescopic device at observation plane *B* will be some $7 \cdot 10^{26}$ km$^2$ in area.  The absolute minimum wait, assuming no star overlap, would be for about $10^{26-11} \cong 10^{15}$ stars to be enclosed; this would take twice $(100/1.2) \cdot 10^{15}$ years, or, well over $10^{17}$ years, a period exceeding the age of the known universe by seven orders of magnitude.

Because Olbers's Conjecture only addresses the issue of why the sky is dark, and not how to illuminate it brightly, and because visible stars are much more radiant than 0, if stars are assumed uniformly distributed in the sky, there is no use to worry about overlapping coverage:  A very sparse coverage would make the sky quite bright.

Using the projecture to predict the night sky, there is no chance that the Olbers Conjecture could yield a sky bright enough to equal that of the surface of the Sun.  The best that might be expected would be a weak but visible glow over 7 log units dimmer than the surface of the sun.  A more precise value is calculated in the SSI posters, below:  Star density averaged over the universe is far lower than 1 per 10 cubic parsecs, as reported in [Robin, et al, 2000].

In the SSI calculations, the average night sky is assumed about 22 – 18 = 4 magnitudes brighter than the darkest possible [Lodriguss, 2003], which amounts to a factor of about $(2.5)^4 \cong 40$.

## The Olbers Analytic Argument

Let's follow through very closely on the Conjecture as usually stated.  In Figure 2, star-filled Gaussian spheres are used to define spherical shells centered on a point of observation such as the Earth.   We postulate a spherically symmetrical, omnidirectional light-detector at the center with total sensitive area *A*.  Without loss of generality, *A* is assumed constant as viewed from any direction in the sky.

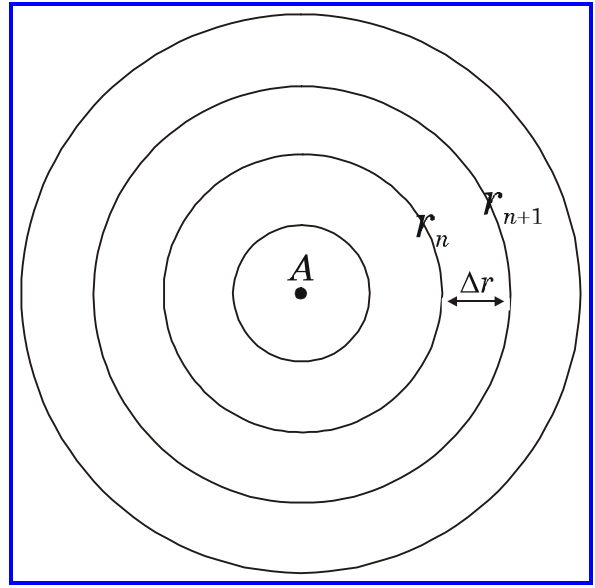

**Figure 2.  Basic construction of the Olbers Conjecture.  Concentric spherical shells of equal thickness $\Delta r$ define the domain of analysis of the stars.**

Assuming that the stars on the average are uniformly distributed in space and have equal luminous flux $L$, let us use $N_n$ to represent the number of stars in the $n$-th shell, between spheres $n$ and $n+1$. The volume of this shell will be given by,

$$V_n = \frac{4}{3}\pi\left[(r_n + \Delta r)^3 - r_n^3\right]. \tag{1}$$

If we assume we are far enough from the center to assign all stars in the $n$-th shell to the same approximate distance $r_n$ from the center, the total light $I_n$ shed on $A$ by the stars in the $n$-th shell will be given by

$$I_n = N_n \cdot L \cdot \frac{A}{4\pi r_n^2}. \tag{2}$$

Taking the average density of stars in space to be $D = N/V$, we may eliminate $N$ between (1) and (2) to write,

$$I_n = D \cdot V_n \cdot L \cdot \frac{A}{4\pi r_n^2} = \frac{1}{3} D \cdot L \cdot A \cdot \frac{1}{r_n^2}\left((r_n + \Delta r)^3 - r_n^3\right). \tag{3}$$

This expression shows the expected behavior of vanishing when $\Delta r \to 0$; it may be rewritten as,

$$I_n = \frac{1}{3} D \cdot L \cdot A \cdot \left(3\Delta r + 3\frac{\Delta r}{r_n} + \frac{\Delta r^3}{r_n^2}\right); \tag{4}$$

so, it is well behaved and approaches a constant as $r_n \to \infty$. Thus, all distant shells contribute about the same amount of light.

Almost all analyses of Olbers stop here. However, let us look at the intensity ratio between two adjacent, increasing shells:

$$\frac{I_{n+1}}{I_n} = \frac{1}{r_{n+1}^2}\left((r_{n+1} + \Delta r)^3 - r_{n+1}^3\right) \bigg/ \frac{1}{r_n^2}\left((r_n + \Delta r)^3 - r_n^3\right). \tag{5}$$

Recalling that $r_{n+1} = r_n + \Delta r$ for all $n$,

$$\frac{I_{n+1}}{I_n} = \frac{r_n^2\left((r_n + 2\Delta r)^3 - (r_n + \Delta r)^3\right)}{(r_n + \Delta r)^2\left((r_n + \Delta r)^3 - r_n^3\right)}; \text{ and, after some algebra,} \tag{6}$$

$$\frac{I_{n+1}}{I_n} = r_n^2 \frac{3r_n^2 + 9r_n\Delta r + 7\Delta r^2}{(r_n + \Delta r)^2\left(3r_n^2 + 3r_n\Delta r + \Delta r^2\right)}. \tag{7}$$

So,

$$\lim_{\Delta r \to 0} \frac{I_{n+1}}{I_n} = 1. \tag{8}$$

This means that if we choose closely-enough spaced shells, the contribution of any adjacent pair of them will be about the same. This is reassuring, but not very important.

Let us multiply out the numerator and denominator of (7) to see what happens if we let the distance go to infinity:

$$\frac{I_{n+1}}{I_n} = \frac{3r_n^4 + 9r_n^3\Delta r + 7r_n^2\Delta r^2}{3r_n^4 + 9r_n^3\Delta r + 10r_n^2\Delta r^2 + 5r_n\Delta r^3 + \Delta r^4}; \text{ and,} \tag{9}$$

the leading coefficients being equal, the numerator and denominator grow at the same rate for $r_n$ large compared with $\Delta r$. This means that,

$$\lim_{r_n \to \infty} \frac{I_{n+1}}{I_n} = 1. \tag{10}$$

So, near the Earth, the closer of any pair of shells contributes a little more than the farther; this is because of the way we have defined $r_n$. And, we find that the relative contribution of adjacent shells grows to 1 with increasing distance. The Olbers Conjecture is that all shells contribute equally, and, so, that this limit indeed should be equal to 1.

Let's now look at the intensity difference between two adjacent shells:

$$\Delta I = I_{n+1} - I_n = (const)\left[\frac{1}{r_{n+1}^2}\left((r_{n+1} + \Delta r)^3 - r_{n+1}^3\right) - \frac{1}{r_n^2}\left((r_n + \Delta r)^3 - r_n^3\right)\right]. \quad (11)$$

Again, because $r_{n+1} = r_n + \Delta r$ for all $n$,

$$\Delta I = (const)\left[\frac{1}{(r_n + \Delta r)^2}\left((r_n + 2\Delta r)^3 - (r_n + \Delta r)^3\right) - \frac{1}{r_n^2}\left((r_n + \Delta r)^3 - r_n^3\right)\right]; \quad (12)$$

$$\Delta I = (const)\left[\frac{-\Delta r^3\left(3r_n^2 + 5r_n\Delta r + \Delta r^2\right)}{r_n^2\left(r_n^2 + 2r_n\Delta r + \Delta r^2\right)}\right] < 0; \text{ but,} \quad (13)$$

$$\lim_{r_n \to \infty} \Delta I = 0. \quad (14)$$

The difference between intensity contributions in adjacent shells is negative, which is to say that the contribution of the more distant shell is less than that of the closer one, as we saw for the ratio above, but this difference decreases to 0 as we look at light from shells farther and farther away.

So, we have again proven the internal consistency of the Olbers Conjecture. However, the Conjecture remains inaccurate, given the reality of the universe as we know it: As is shown in the SSI posters below, there aren't enough of stars.

# SSI-31 Poster Session: If Olbers Had Had a Telescope

## Heinrich Wilhelm Olbers (1823) — 1

**Conjecture:** If stars on average are distributed uniformly, and the universe is infinite, the sky at night should be bright.

$N_n$ = number of stars in the volume $V_n$ between spherical shells $n$ and $n+1$.

Then, $V_n = (4\pi/3)[(r_n + \Delta r)^3 - r_n^3]$.   If the average total luminous flux of a star is $\langle L \rangle$, the total light $I_n$ shed on a detector of area $A$ by the $n$-th shell is, $I_n = N_n \langle L \rangle A/(4\pi r_n^2)$

If the average density of stars in the universe is $\langle D \rangle = \langle N \rangle / V$,

$$I_n = (1/3)\langle D \rangle \langle L \rangle A (1/r_n^2)[(r_n + \Delta r)^3 - r_n^3].$$

For all $n$, we know that $r_{n+1} = r_n + \Delta r$; so,

$$I_{n+1}/I_n = r_n^2 \frac{3r_n^2 + 9r_n\Delta r + 7\Delta r^2}{(r_n + \Delta r)^2 (3r_n^2 + 3r_n\Delta r + \Delta r^2)}.$$

Then, $\lim_{r_n \to \infty}(I_{n+1}/I_n) = \lim_{r_n \to \infty}(r_n^4/r_n^4) = 1$.

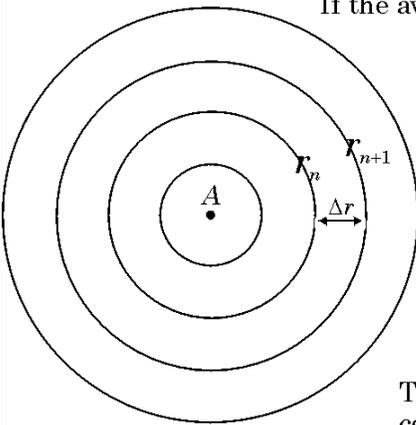

This shows that the Conjecture is correct: All shells contribute equally, for sufficiently large distances.

See http://arXiv.org/pdf/physics/0104067 for details of these posters.

## A Telescopic Device   2

**Earth** → Direction of Extension → Cap of cylinder

Light-gathering Cone

1 parsec

An imaginary cylinder of constant, 1-parsec diameter, elongates itself almost at speed $c$, gaining 3 parsecs in length, or 2.4 parsec$^3$ in volume, per 10 years. The cylinder is opaque, with completely absorbing walls and end-cap.  Stars intercepted may drift in and out of the cylinder.  The cylinder contains virtual, longitudinal optic fibers which convey all star-images to the plane of the entrance of the light-gathering cone, at magnification of 1.  This removes the ordinary square-law dependence of observed star area on distance.  An observer on Earth may determine when that plane is uniformly illuminated at the (average) surface brightness of a star. Taking the **uncorrected** density in the universe of 0.1 star/parsec$^3$, the cylinder then averages about 2.4 new stars per century.  Let's say the observer then sees about 1 new star per century.

## Telescopic Sampling of the Universe   3

Assuming the average star to be $10^6$ km in diameter; the projected area of that star then would be about $8 \cdot 10^{11}$ km$^2$.
A parsec is about $3 \cdot 10^{13}$ km; the telescopic end-cap then has an area of $7 \cdot 10^{26}$ km$^2$. So, it would require the area of $10^{15}$ average stars to fill the cone field of view, bringing the intensity up to that of the surface of an average star.

We are interested only in low densities of stars, so we shall assume that they rarely or never overlap (obscure one another) in the telescoping cylinder.

From Wyszecki & Stiles (1982), detection threshold for light in a large patch of sky will be taken as $(1/360) \cdot 10^{-10} \cong 3 \cdot 10^{-12}$ of the irradiance by the Sun on the Earth's surface.  So, light from empty sky populated per unit area less than $10^{-12}$ with stars will not be noticed.  Light considerably above this threshold will brighten visibly with increased area coverage by stars, and there will be a smooth transition between these regimes.

Because of the visual detection threshold, we conclude that no light at all can be seen at the entrance of the light-gathering cone until the cylinder of the telescopic device contains an average of about $10^{15} \cdot 3 \cdot 10^{-12} \cong 3000$ stars.

Lodriguss (2003) allows us to estimate that a very "dark" sky is about 4 stellar magnitudes brighter than the darkest known sky; and, a deep-sky study by Robin, *et al* (2000) tells us that stars in the halo of a galaxy number about $1.6 \cdot 10^{-4}$/parsec$^3$.

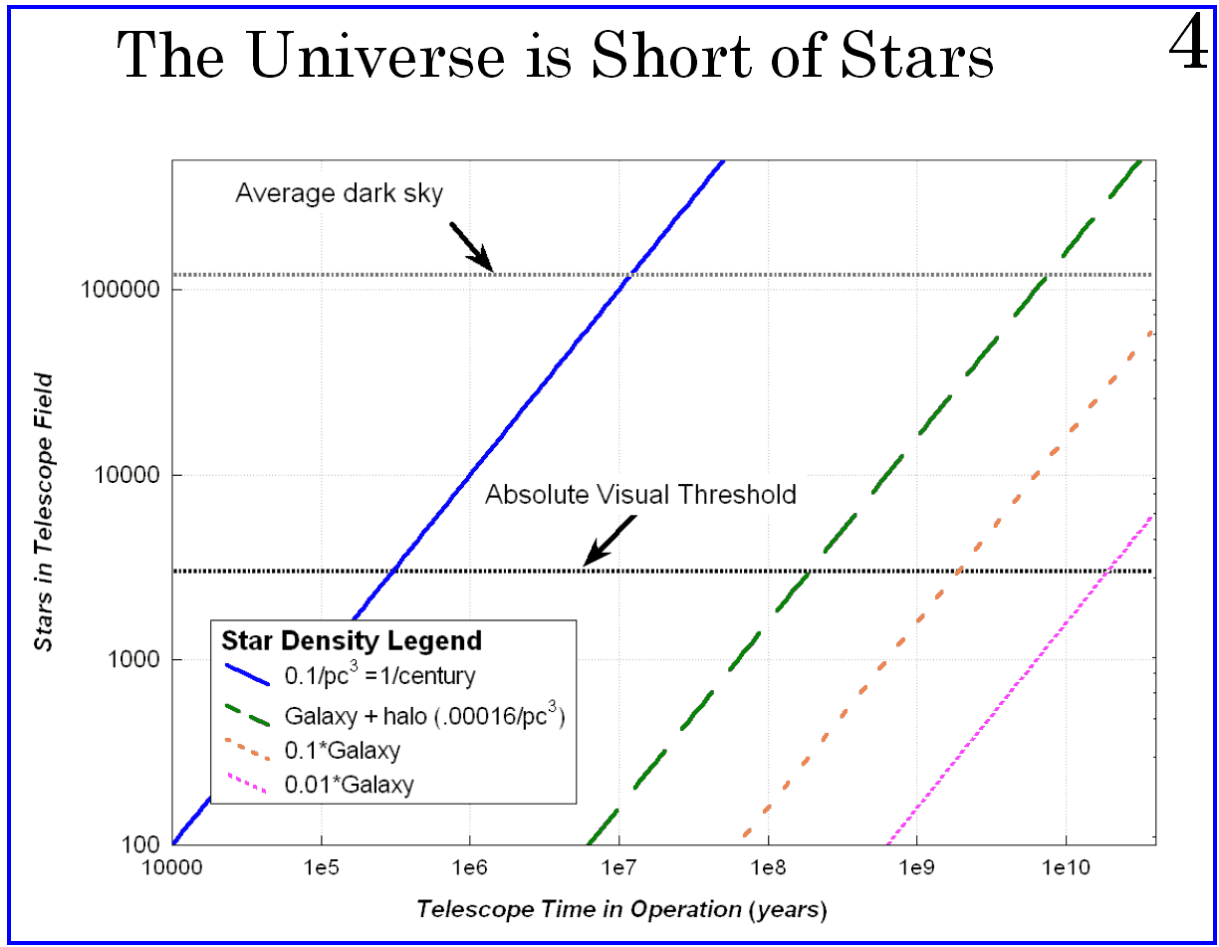